\documentclass[sigconf,screen,natbib=false]{acmart}

\usepackage{tikz}
\usepackage{multirow}
\usepackage{pifont}
\usepackage[breakable]{tcolorbox}
\usepackage{fvextra}
\usepackage{subcaption}

\definecolor{lightblack}{gray}{0.3} % 0.0 is black, 1.0 is white. 0.3 is dark gray/light black.

\setcopyright{acmlicensed}
\copyrightyear{2026}
\acmYear{2026}
\acmDOI{XXXXXXX.XXXXXXX}
\acmConference[]{}{}{}
\acmISBN{978-1-4503-XXXX-X/2026/10}

\RequirePackage[
  datamodel=acmdatamodel,
  style=acmnumeric,
  ]{biblatex}

\begin{document}

\title{Graded-Relevance Composed Multimodal Retrieval for E-commerce Visual Search at Scale}
\author{Anubhav Gupta}
\affiliation{%
  \institution{Walmart Global Tech}
  \city{Bangalore}
  \country{India}}
\email{anubhav.gupta0@walmart.com}

\author{Hrushikesh Mohapatra}
\affiliation{%
  \institution{Walmart Global Tech}
  \city{Bangalore}
  \country{India}}
\email{Hrushikesh.Mohapatra@walmart.com}

\author{Prijith Chandra}
\affiliation{%
  \institution{Walmart Global Tech}
  \city{Sunnyvale}
  \country{United States}}

\author{Asish Mohapatra}
\affiliation{%
  \institution{Walmart Global Tech}
  \city{Bangalore}
  \country{India}}

\author{Anuj Garg}
\affiliation{%
  \institution{Walmart Global Tech}
  \city{Bangalore}
  \country{India}}

\author{Arvind Maan}
\affiliation{%
  \institution{Walmart Global Tech}
  \city{Bangalore}
  \country{India}}

\author{Sudip Datta}
\affiliation{%
  \institution{Walmart Global Tech}
  \city{Bangalore}
  \country{India}}

\author{Venkat Bulusu}
\affiliation{%
  \institution{Walmart Global Tech}
  \city{Bangalore}
  \country{India}}

\author{Sitesh Kumar Jalan}
\affiliation{%
  \institution{Walmart Global Tech}
  \city{Bangalore}
  \country{India}}

\renewcommand{\shortauthors}{Gupta et al.}




\begin{abstract}
  Visual search on large e-commerce catalogs must serve both ``similarity'' queries that ask for items resembling an uploaded image and ``modifier'' queries that comprise an image and text describing a desired modification (e.g.\ a color change or style swap). The latter is the setting known as \emph{composed image retrieval} (CIR). Existing CIR methods, however, treat relevance as binary and train on triplets with a single positive target---a poor fit for real catalogs where many candidates partially satisfy a user query and ranking across that partial-match spectrum drives the customer experience. We propose a methodology for training CIR retrievers on graded relevance, consisting of: (i) a VLM to curate training data, generating both queries (object detection + modifier synthesis) and 4-level relevance labels without manual annotation, (ii) an iterative relevance-feedback loop that expands the training set by mining hard negatives from the in-training retriever, and (iii) a hierarchy-aware angular objective to train the retriever directly on the graded labels rather than collapsing them to a binary split. We call this methodology GradCIR and instantiate it on a PaliGemma2 bi-encoder trained on 3.5M graded pairs curated from raw Walmart catalog data. A controlled graded-vs-binary ablation isolates the supervision granularity and shows lift of 4.9\%--5.9\% in NDCG@10. The same recipe applied to other multimodal encoders lifts early-fusion backbones by up to 8.5\% NDCG@10. On the public FashionIQ benchmark, GradCIR (applied to PaliGemma2) reaches 0.6703 average recall when fine-tuned, slightly ahead of the strongest peer-reviewed supervised baseline we compare against, and matching or exceeding all published CLIP-L-class zero-shot CIR methods. The system is deployed in production at Walmart, where it's serving live visual-search user traffic.

\end{abstract}

\maketitle

\section{Introduction}
\begin{figure}[!t]
    \vspace{0.5cm}
    \centering
    \begin{subfigure}[t]{0.48\columnwidth}
        \centering
        \includegraphics[width=\textwidth]{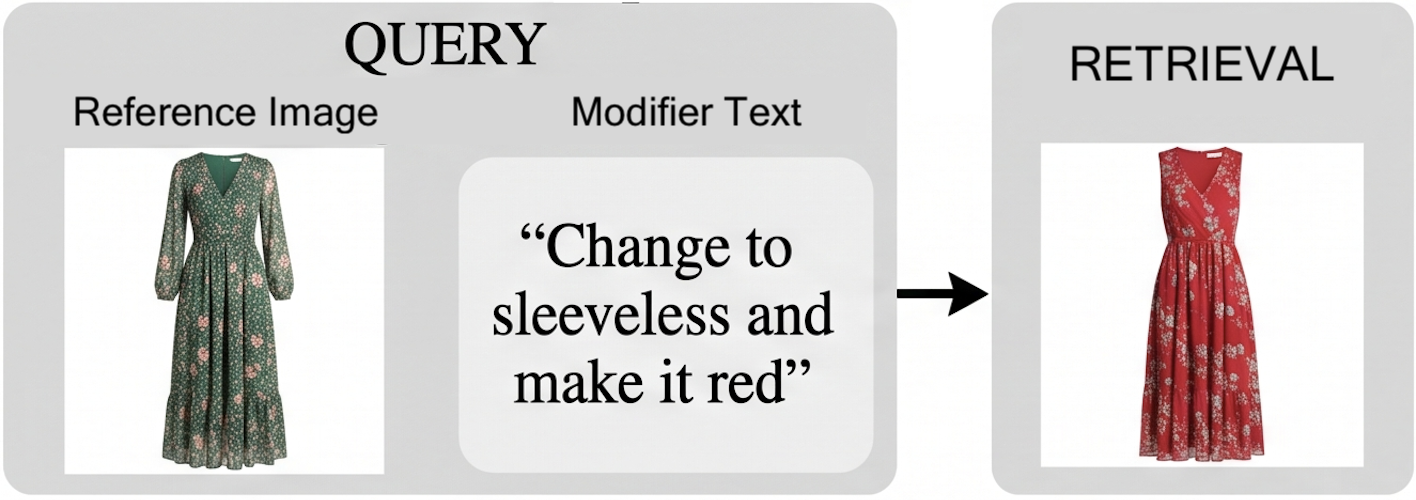}
        \caption{\emph{Exact:} all visual attributes align.}
    \end{subfigure}
    \hfill
    \begin{subfigure}[t]{0.48\columnwidth}
        \centering
        \includegraphics[width=\textwidth]{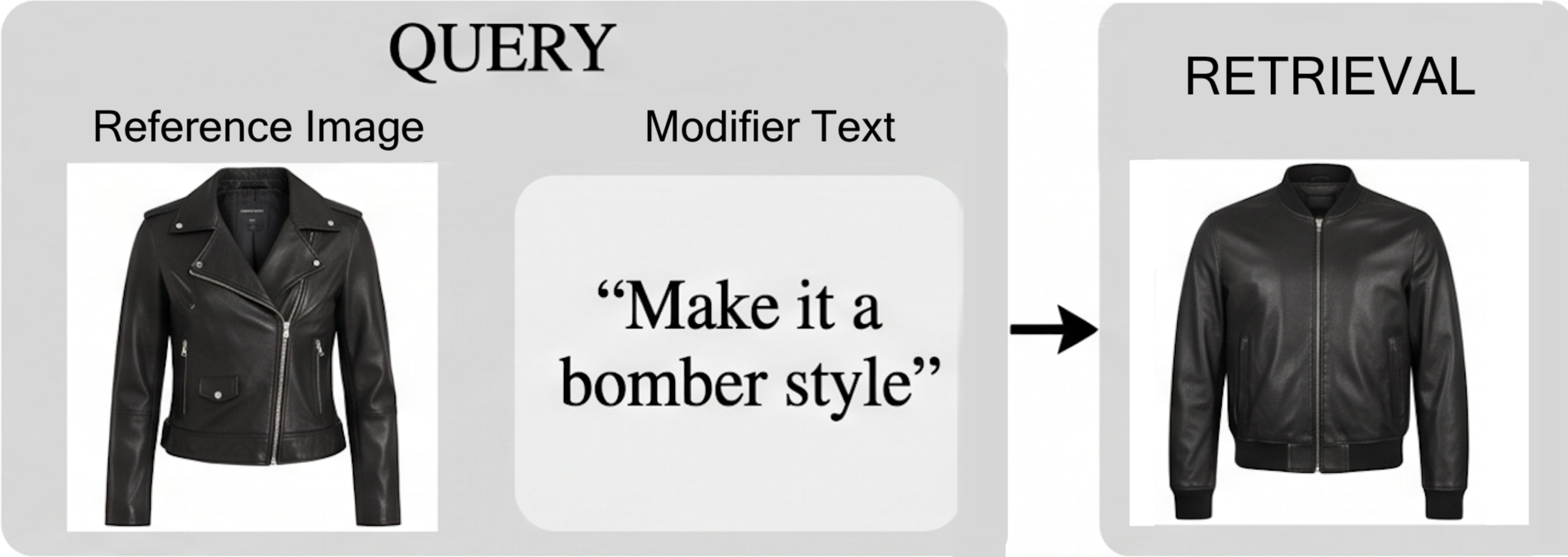}
        \caption{\emph{Near Exact:} style modifier; core attributes (colour, material) retained but not identical.}
    \end{subfigure}

    \vspace{0.2cm}

    \begin{subfigure}[t]{0.48\columnwidth}
        \centering
        \includegraphics[width=\textwidth]{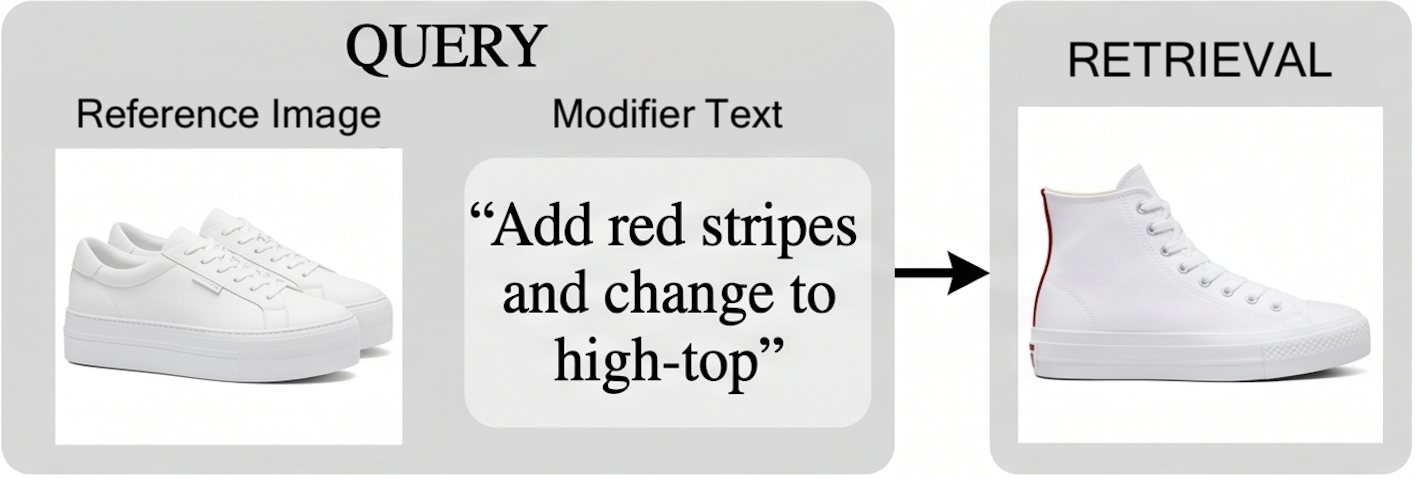}
        \caption{\emph{Partial Match:} category and partial attribute (high-top) match, but a major feature (red stripes) is missing.}
    \end{subfigure}
    \hfill
    \begin{subfigure}[t]{0.48\columnwidth}
        \centering
        \includegraphics[width=\textwidth]{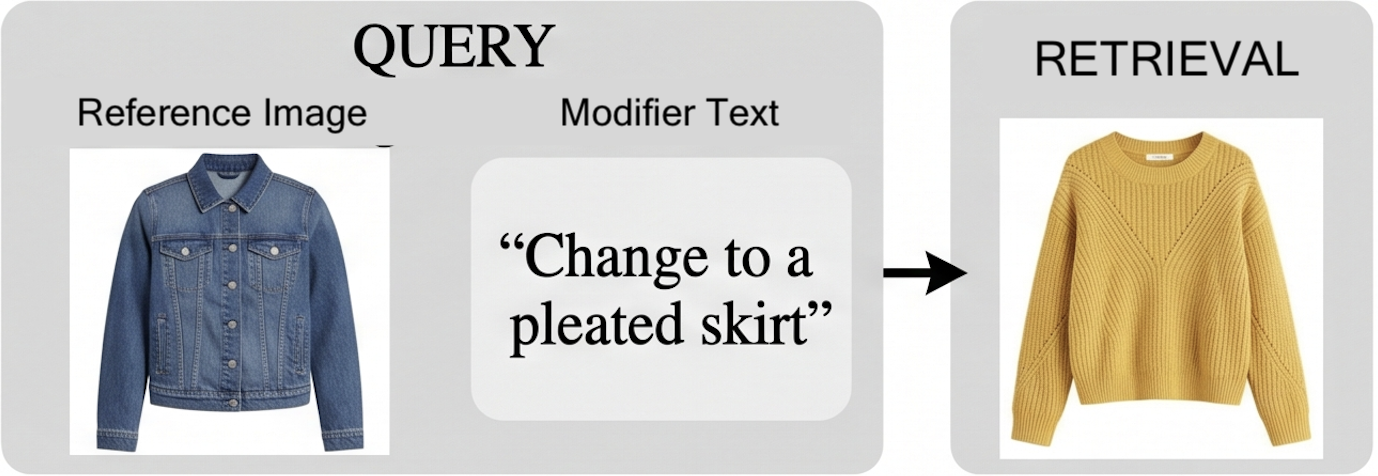}
        \caption{\emph{Irrelevant:} different product type and attributes.}
    \end{subfigure}
    \caption{An example of graded CIR relevance labeling over a 4-level ordinal scale.}
    \label{fig:graded_relevance}
\end{figure}

Visual search is at the core of many e-commerce platforms today, allowing customers to search for products using a multimodal query consisting of an image and the accompanying text. A typical visual search system consists of (1) a \emph{retriever} that takes the user query and returns a set of candidate items from the catalog, and (2) a \emph{ranker} that orders the retrieved candidates by relevance. This paper focuses on the retrieval component, which is typically realised as deep metric learning over high-dimensional embeddings: relevant (query, item) pairs are placed closer in the embedding space than irrelevant ones, and approximate nearest-neighbour (ANN) search retrieves the top-$K$ candidates at inference time.

Many user queries, especially for categories like Fashion, have a visual aspect (colour, appearance, design) which can be difficult to express in words. Most existing systems, however, tackle only the limited use-case of ``shop similar items''~\cite{hadi2015buy,du2022amazon,zhu2024bringing}. Building a robust visual search system at production scale poses several unique technical challenges. The first is \emph{domain shift}: query images and product images come from different distributions, as catalog images are usually clean and high-resolution, whereas user-uploaded images contain noise, clutter, and irrelevant background. The second is \emph{catalog scale}: Walmart's catalog contains hundreds of millions of items, and retrieving tens of highly relevant items with low latency is critical to a good customer experience. The third is \emph{relevance gradation} (Figure~\ref{fig:graded_relevance}): in a real catalog, candidates are rarely cleanly ``relevant or irrelevant''. For a query of a grey velvet 3-seater sofa, an identical product is an exact match; the same sofa in beige is a strong near match; a generic grey 3-seater is a partial match; a velvet armchair is irrelevant. Collapsing these four cases into a binary labeling scheme discards the information that makes a retrieval ranking useful.

The dominant line of work on composed image retrieval (CIR)~\cite{baldrati2022clip4cir,liu2024bi,bai2024sentence,feng2024spn4cir,zhang2024gme,huynh2025collm,zhang2024magiclens} optimizes for the binary case: triplets of (reference image, modifier text, target image) where exactly one target is correct. This works well for the academic CIR benchmarks but is a poor fit for production retrieval over a real catalog, where the ranking quality across a partial-match spectrum is what drives customer engagement. This paper presents \emph{GradCIR}, a methodology for training composed-multimodal retrievers on graded ordinal relevance. The methodology is backbone-agnostic: in the experiments we instantiate it on PaliGemma2 (referred to as \emph{GradCIR-PG2} subsequently) and additionally apply the same recipe to four other contemporary multimodal encoders to confirm that the gains are properties of the methodology rather than of any particular backbone. The main contributions of the paper are:

\begin{enumerate}
    \item A \textbf{graded-relevance formulation of CIR} over a 4-level ordinal scale, trained with a hierarchy-aware angular objective. To our knowledge, all prior CIR work collapses relevance to a single positive target. A controlled ablation (\S\ref{sec:ablation_graded}) shows that finer grading produces monotonically higher NDCG with backbone, data, and loss held fixed.
    \item A \textbf{manual-annotation-free, VLM-curated data pipeline} that drives both query generation and graded relevance labeling end-to-end. This is coupled with a relevance-feedback loop that grows the training set by mining hard negatives from the in-training retriever. The pipeline produces 3.5M graded pairs on the Walmart catalog.
    \item \textbf{Competitive public-benchmark results.} On FashionIQ, GradCIR-PG2 reaches Avg(R@10,R@50)=0.6703, slightly ahead of the strongest peer-reviewed supervised baseline we compare against (SPN4CIR, 0.6641), in supervised setting. In zero-shot setting, it matches or exceeds published CLIP-L-class zero-shot CIR methods without any CIR-triplet supervision.
\end{enumerate}

The remainder of the paper is organised as follows. Section~\ref{sec:related_work} discusses related work on visual search, VLM-based embeddings, and composed (multimodal) retrieval. Section~\ref{sec:methodology} describes our proposed data-creation pipeline, model architecture, and graded-relevance training objective. Section~\ref{sec:experiments} reports results on public benchmarks and our internal Walmart Visual Search Test set, followed by ablations and deployment details. Section~\ref{sec:scope_limitations} discusses scope and limitations of our work. Section~\ref{sec:conclusion} concludes with a summary and future directions.

% End-to-end architecture diagram for GradCIR
% Requires: \usepackage{tikz} and the libraries loaded below
% Image files expected in: ../figures/arch_figs/

\usetikzlibrary{
  positioning,
  arrows.meta,
  shapes.geometric,
  shapes.symbols,
  fit,
  calc,
  backgrounds,
  decorations.pathreplacing
}

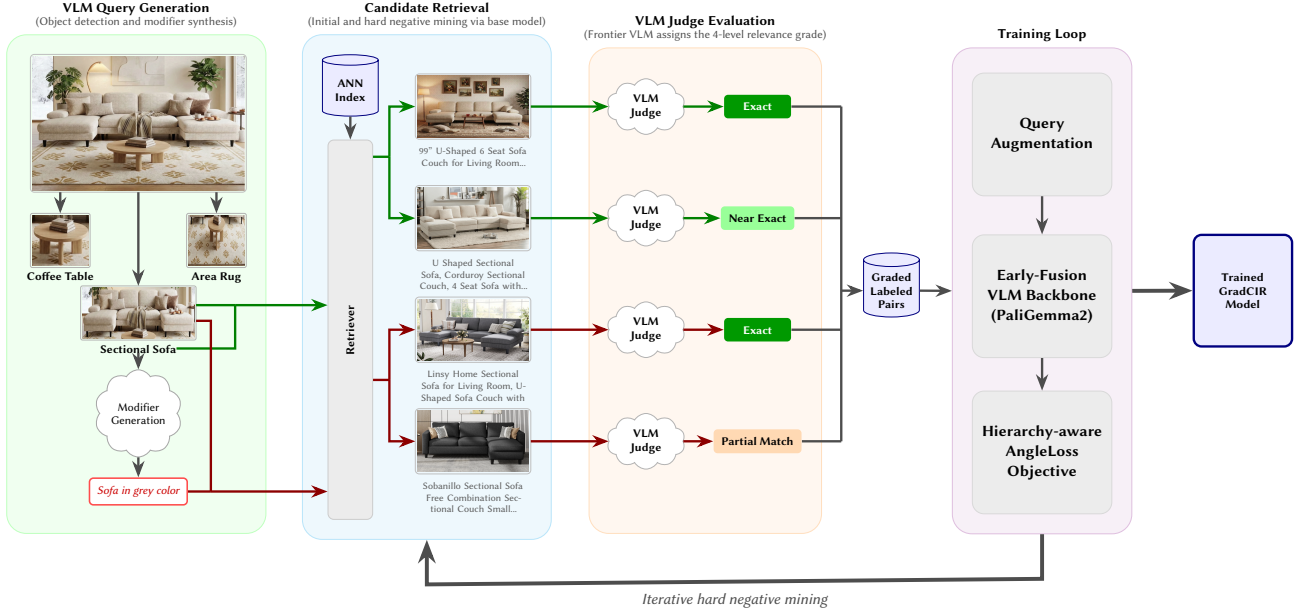
\begin{figure*}[t]
\centering
\resizebox{0.96 \textwidth}{!}{%
\begin{tikzpicture}[
    >=Stealth,
    font=\sffamily\small,
    % --- Node styles ---
    imgnode/.style={inner sep=1pt, outer sep=0pt, draw=gray!50, rounded corners=2pt, line width=0.5pt},
    cropnode/.style={inner sep=1pt, outer sep=0pt, draw=gray!60, rounded corners=1pt, line width=0.5pt},
    croplabel/.style={font=\sffamily\scriptsize\bfseries, text=black, inner sep=0pt},
    cloudstyle/.style={cloud, cloud puffs=10, cloud puff arc=120, draw=gray!70, fill=white,
                       inner sep=0pt, font=\sffamily\scriptsize, align=center,
                       minimum width=1.4cm, minimum height=0.2cm},
    modbox/.style={draw=red!70, fill=white, rounded corners=2pt, font=\sffamily\scriptsize\itshape,
                   inner sep=4pt, text=red!80!black, line width=0.8pt},
    retrieverbox/.style={fill=gray!15, draw=gray!40, rounded corners=3pt, minimum width=0.8cm,
                         font=\sffamily\scriptsize\bfseries, align=center},
    cylinderstyle/.style={cylinder, shape border rotate=90, draw=blue!50!black, fill=blue!8,
                          aspect=0.25, minimum height=1.1cm, minimum width=1.0cm,
                          font=\sffamily\scriptsize\bfseries, align=center},
    vlmjudge/.style={cloud, cloud puffs=10, cloud puff arc=120, draw=gray!60, fill=white,
                     inner sep=1pt, font=\sffamily\scriptsize\bfseries, align=center,
                     minimum width=1.4cm, minimum height=0.9cm},
    exactlabel/.style={fill=green!60!black, text=white, rounded corners=2pt,
                       font=\sffamily\scriptsize\bfseries, inner sep=4pt, minimum width=1.2cm},
    nearexactlabel/.style={fill=green!40, text=black, rounded corners=2pt,
                           font=\sffamily\scriptsize\bfseries, inner sep=4pt, minimum width=1.2cm},
    partiallabel/.style={fill=orange!30, text=black, rounded corners=2pt,
                         font=\sffamily\scriptsize\bfseries, inner sep=4pt, minimum width=1.2cm},
    grayblock/.style={fill=gray!15, draw=gray!30, rounded corners=6pt, minimum width=2.5cm,
                      minimum height=2.2cm, font=\sffamily\small\bfseries, align=center},
    outputlabel/.style={draw=blue!50!black, fill=blue!8, rounded corners=4pt, line width=1.2pt,
                        font=\sffamily\scriptsize\bfseries, align=center, inner sep=8pt,
                        text width=1.2cm, minimum height=2.0cm},
    arrowstyle/.style={->, green!50!black, line width=1.2pt},
    arrowstylemod/.style={->, red!55!black, line width=1.2pt},
    arrowblack/.style={->, black!70, line width=1.2pt},
    arrowblackthick/.style={->, black!70, line width=2.0pt},
    arrowstylethick/.style={->, green!50!black, line width=2.0pt},
    titletext/.style={font=\sffamily\tiny, text width=2.5cm, align=center, text=gray!70!black},
    % Zone caption style — above zone, black, larger font
    zonecaption/.style={font=\sffamily\footnotesize\bfseries, text=black, anchor=south, align=center},
]

% Subtitle command for zone captions
\newcommand{\zonesubtitle}[1]{{\fontseries{m}\selectfont\sffamily\fontsize{6}{7}\selectfont\textcolor{gray!60!black}{#1}}}

% =====================================================================
% Shared vertical bounds for all zones (must exceed all content extents)
% Catalog top ~10.2, retriever bottom ~-0.6 → use 10.3 / -0.7
% =====================================================================
\pgfmathsetmacro{\ztop}{9.2}
\pgfmathsetmacro{\zbot}{1.2}

% =====================================================================
% SECTION 1: Query Generation (left green zone, x = 0..5)
% =====================================================================

% Catalog image — top of left zone (y=8.8 to align zone top with zone 2)
\node[imgnode] (catalog) at (2.2, 8.2) {%
  \includegraphics[width=3.8cm, height=2.4cm]{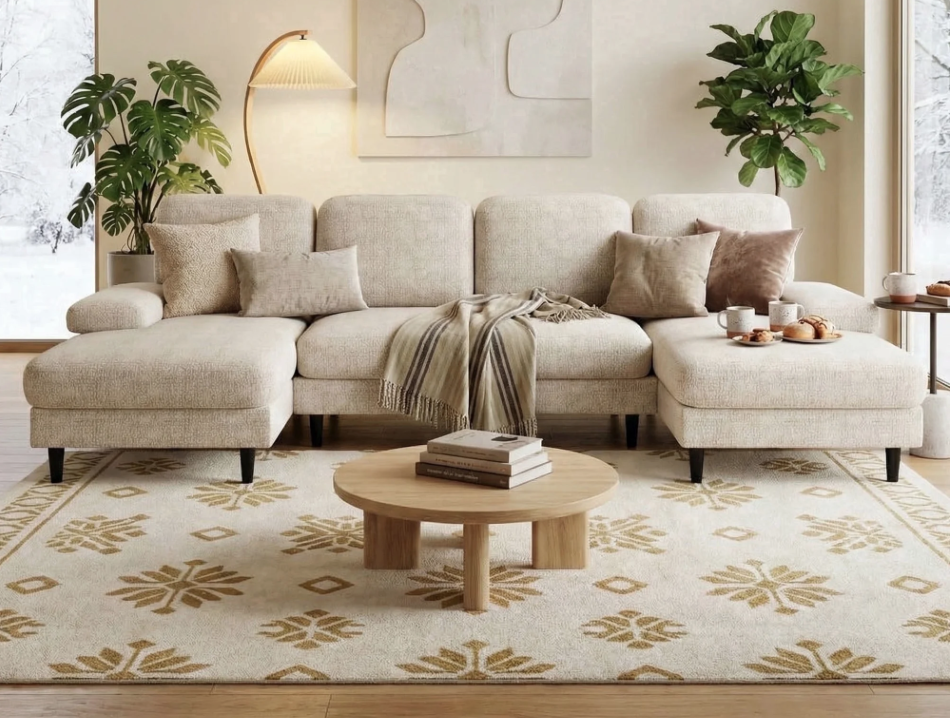}%
};

% --- Detected crops: Coffee Table and Area Rug on sides, Sofa centered below ---

% Crop: Coffee Table — upper left
\node[cropnode] (crop_table) at (0.8, 6.1) {%
  \includegraphics[width=1.0cm, height=0.9cm]{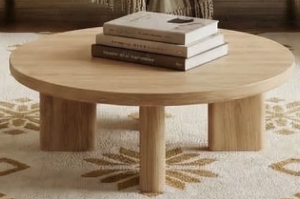}%
};
\node[croplabel, below=2pt of crop_table] (label_table) {Coffee Table};

% Crop: Area Rug — upper right
\node[cropnode] (crop_rug) at (3.6, 6.1) {%
  \includegraphics[width=1.0cm, height=0.9cm]{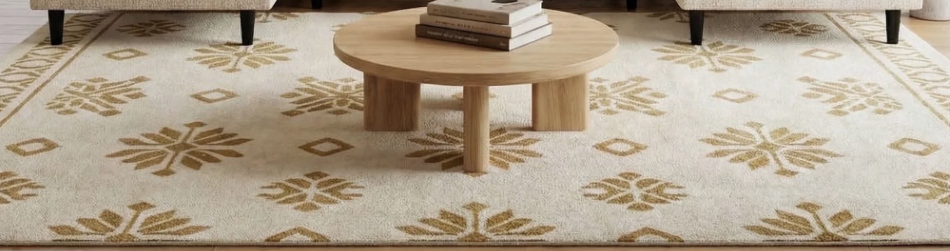}%
};
\node[croplabel, below=2pt of crop_rug] (label_rug) {Area Rug};

% Crop: Sectional Sofa — center, below Coffee Table / Area Rug
\node[cropnode] (crop_sofa) at (2.2, 4.8) {%
  \includegraphics[width=2.0cm, height=0.9cm]{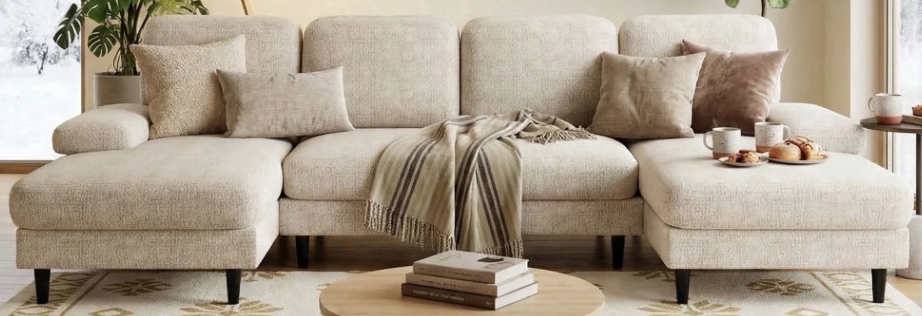}%
};
\node[croplabel, below=2pt of crop_sofa] (label_sofa) {Sectional Sofa};

% --- Modifier pipeline (vertically centered below sofa) ---

% Modifier Generation cloud — centered below sofa label
\node[cloudstyle] (modifier_gen) at (2.2, 3.0) {Modifier\\Generation};

% Modifier text box — below modifier generation
\node[modbox] (modifier_text) at (2.2, 1.6) {Sofa in grey color};

% Zone 1 background (explicit y-spacers for consistent height)
\coordinate (z1top) at (2.2, \ztop);
\coordinate (z1bot) at (2.2, \zbot);
\begin{scope}[on background layer]
  \node[fit=(catalog)(crop_table)(label_table)(crop_rug)(label_rug)(modifier_gen)(modifier_text)(z1top)(z1bot),
        fill=green!6, draw=green!25, rounded corners=10pt,
        inner sep=10pt] (querybg) {};
\end{scope}
% Zone 1 caption
\node[zonecaption] at (querybg.north) {VLM Query Generation\\[-1pt]
  \zonesubtitle{(Object detection and modifier synthesis)}};

% =====================================================================
% SECTION 2: ANN Index + Retriever (x ~ 6..7)
% =====================================================================

% ANN Index cylinder — above retriever
\node[cylinderstyle] (ann_index) at (6.0, 8.8) {ANN\\Index};

% Retriever bar — tall, spanning all 4 rows
\node[retrieverbox, minimum height=6.8cm] (retriever) at (6.0, 4.5) {\rotatebox{90}{Retriever}};

% =====================================================================
% SECTION 3: Retrieved Items + VLM Judge + Labels (x ~ 8..15)
% =====================================================================

% --- Row 1: Exact (top, y=8.5) ---
\node[imgnode] (ret_img1) at (8.2, 8.5) {%
  \includegraphics[width=2.0cm, height=1.1cm]{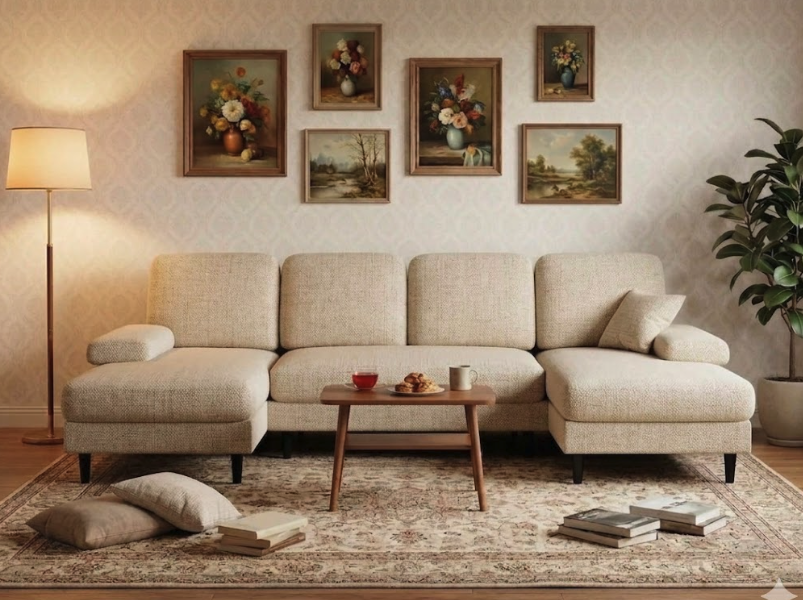}%
};
\node[titletext, below=1pt of ret_img1] (ret_title1) {99'' U-Shaped 6 Seat Sofa Couch for Living Room...};
\node[vlmjudge] (vlm1) at (11.3, 8.5) {VLM\\Judge};
\node[exactlabel] (label_exact1) at (13.3, 8.5) {Exact};

% --- Row 2: Near Exact (y=6.0) ---
\node[imgnode] (ret_img2) at (8.2, 6.5) {%
  \includegraphics[width=2.0cm, height=1.1cm]{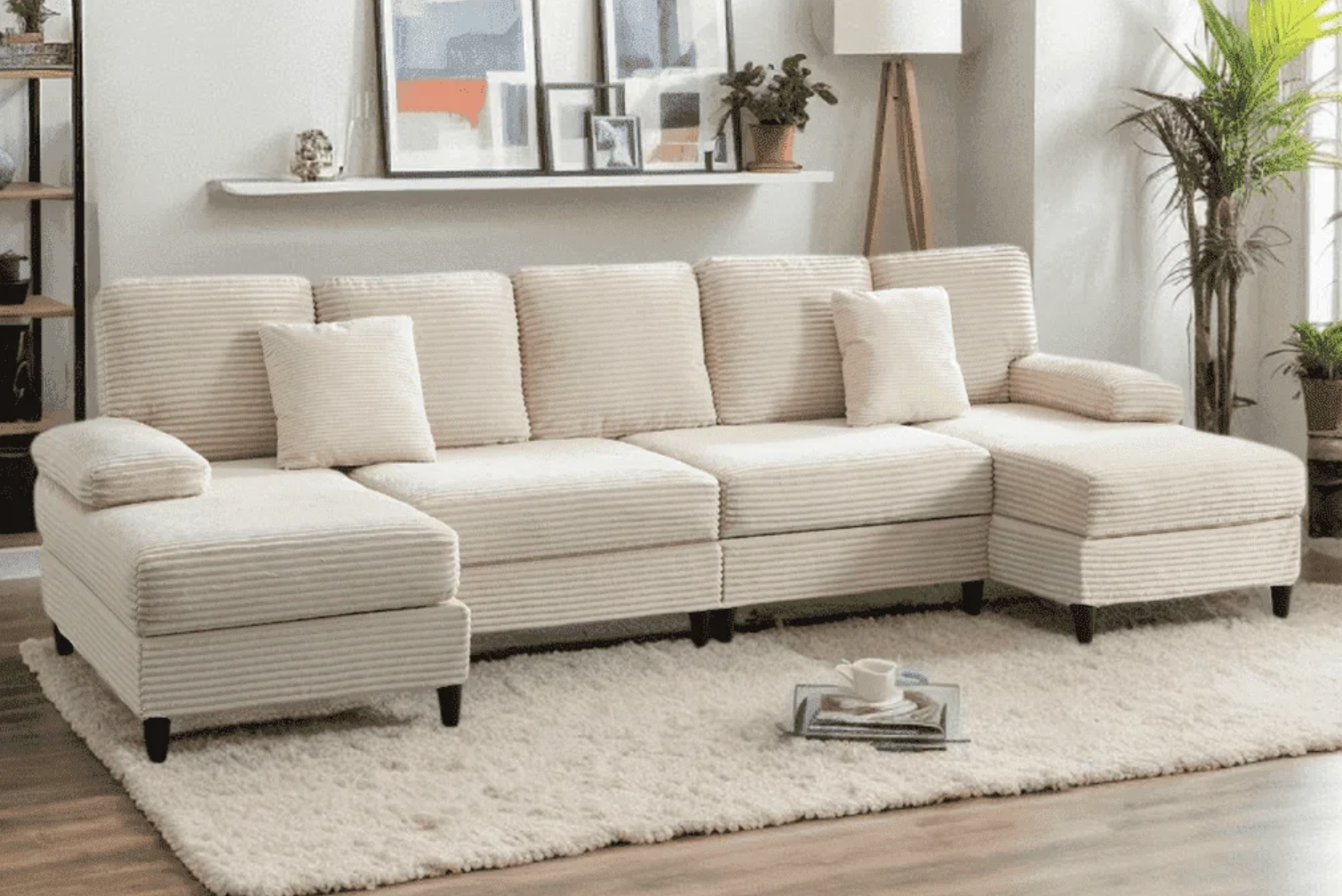}%
};
\node[titletext, below=1pt of ret_img2] (ret_title2) {U Shaped Sectional Sofa, Corduroy Sectional Couch, 4 Seat Sofa with...};
\node[vlmjudge] (vlm2) at (11.3, 6.5) {VLM\\Judge};
\node[nearexactlabel] (label_nearexact) at (13.3, 6.5) {Near Exact};

% --- Row 3: Exact (modifier query, y=3.5) ---
\node[imgnode] (ret_img3) at (8.2, 4.5) {%
  \includegraphics[width=2.0cm, height=1.1cm]{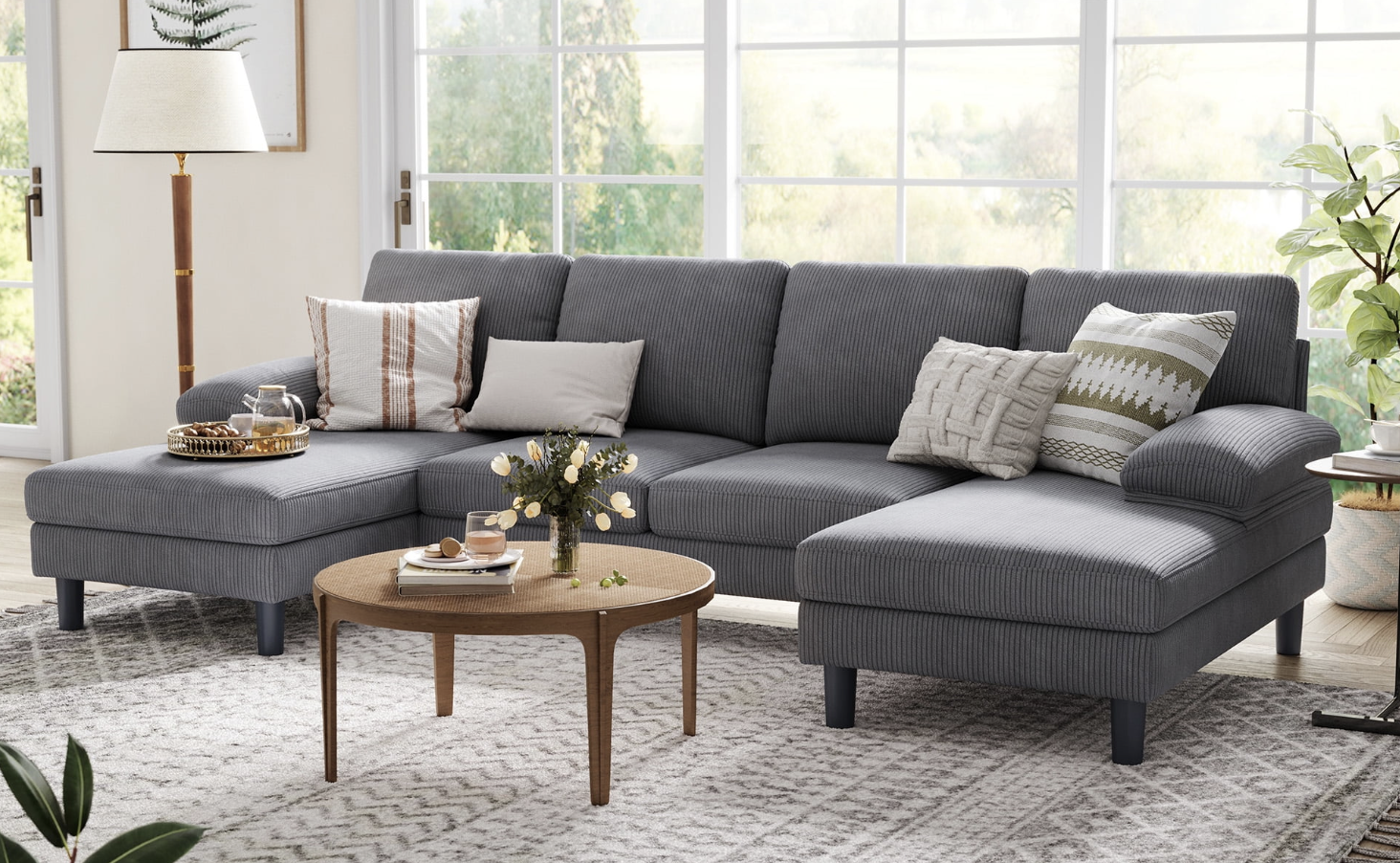}%
};
\node[titletext, below=1pt of ret_img3] (ret_title3) {Linsy Home Sectional Sofa for Living Room, U-Shaped Sofa Couch with};
\node[vlmjudge] (vlm3) at (11.3, 4.5) {VLM\\Judge};
\node[exactlabel] (label_exact2) at (13.3, 4.5) {Exact};

% --- Row 4: Partial Match (bottom, y=1.0) ---
\node[imgnode] (ret_img4) at (8.2, 2.5) {%
  \includegraphics[width=2.0cm, height=1.1cm]{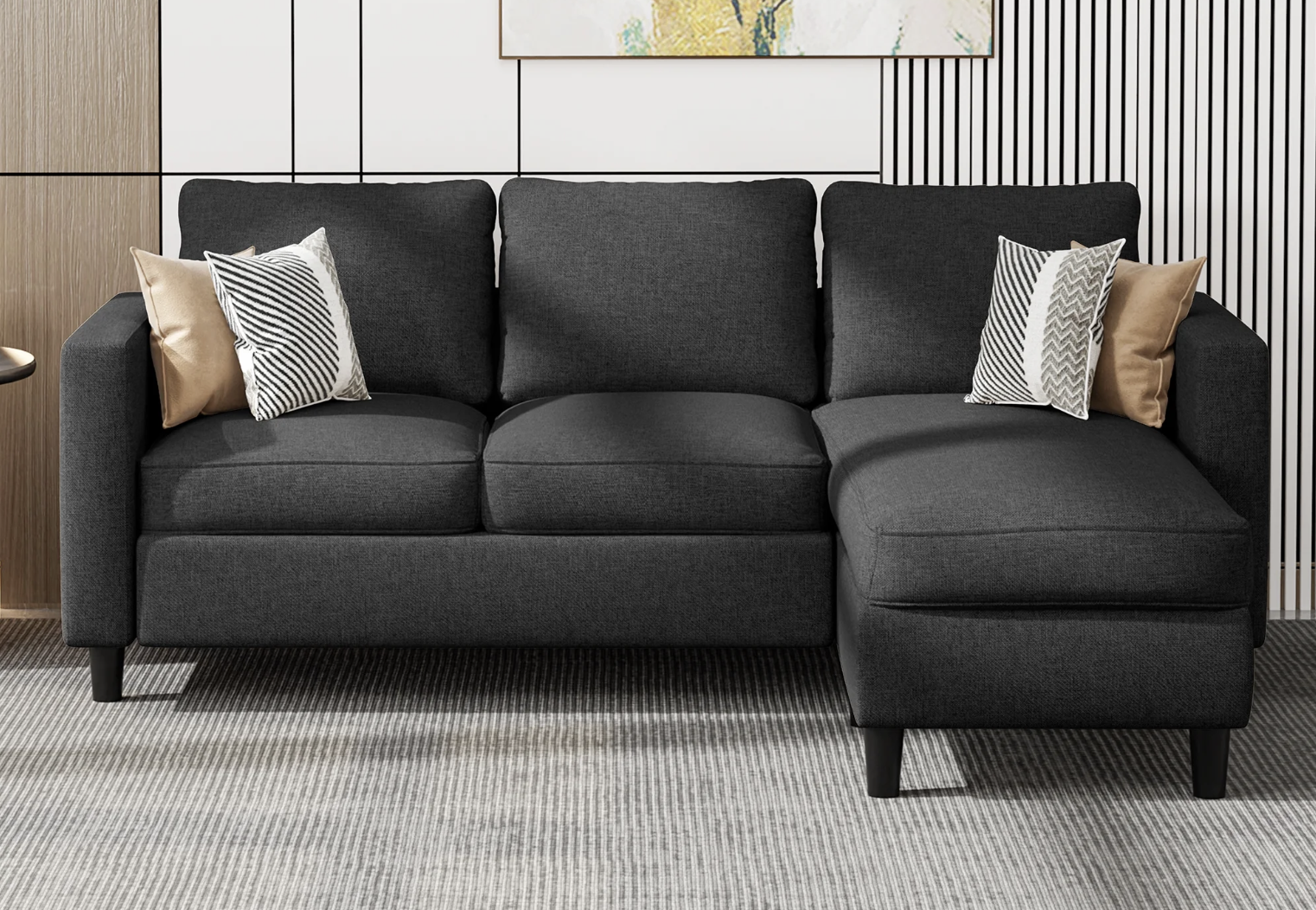}%
};
\node[titletext, below=1pt of ret_img4] (ret_title4) {Sobanillo Sectional Sofa Free Combination Sectional Couch Small...};
\node[vlmjudge] (vlm4) at (11.3, 2.5) {VLM\\Judge};
\node[partiallabel] (label_partial) at (13.3, 2.5) {Partial Match};

% Zone 2 background (explicit y-spacers for consistent height)
\coordinate (z2top) at (6.0, \ztop);
\coordinate (z2bot) at (6.0, \zbot);
\coordinate (z2right) at (9.25, 5.0);  % right-edge limiter: 0.2cm inward from title text edge
\begin{scope}[on background layer]
  \node[fit=(ann_index)(retriever)(ret_img1)(ret_img4)(z2right)(z2top)(z2bot),
        fill=cyan!6, draw=cyan!25, rounded corners=10pt,
        inner sep=10pt] (retrievalbg) {};
\end{scope}
% Zone 2 caption
\node[zonecaption] at (retrievalbg.north) {Candidate Retrieval\\[-1pt]
  \zonesubtitle{(Initial and hard negative mining via base model)}};

% Zone 3 background (explicit y-spacers for consistent height)
\coordinate (z3top) at (12.3, \ztop);
\coordinate (z3bot) at (12.3, \zbot);
\begin{scope}[on background layer]
  \node[fit=(vlm1)(label_exact1)(vlm4)(label_partial)(z3top)(z3bot),
        fill=orange!6, draw=orange!25, rounded corners=10pt,
        inner sep=10pt] (labelingbg) {};
\end{scope}
% Zone 3 caption
\node[zonecaption] at (labelingbg.north) {VLM Judge Evaluation\\[-1pt]
  \zonesubtitle{(Frontier VLM assigns the 4-level relevance grade)}};

% =====================================================================
% SECTION 4: Graded Labeled Pairs (x ~ 16)
% =====================================================================

\node[cylinderstyle] (graded_db) at (15.7, 5.2) {Graded\\Labeled\\Pairs};

% =====================================================================
% SECTION 5: Model Training (right green zone, x ~ 18..24)
% =====================================================================

% Query Augmentation
\node[grayblock] (query_aug) at (18.4, 8.0) {Query\\Augmentation};

% Early-Fusion VLM Backbone
\node[grayblock] (vlm_backbone) at (18.4, 5.1) {Early-Fusion\\VLM Backbone\\(PaliGemma2)};

% Hierarchy-aware AngleLoss Objective
\node[grayblock] (angleloss) at (18.4, 2.3) {Hierarchy-aware\\AngleLoss\\Objective};

% Arrows between training blocks (vertical flow)
\draw[arrowblack] (query_aug.south) -- (vlm_backbone.north);
\draw[arrowblack] (vlm_backbone.south) -- (angleloss.north);

% Trained GradCIR Model output
\node[outputlabel] (output_model) at (22.0, 5.2) {Trained\\GradCIR\\Model};

% Zone 4 background (explicit y-spacers for consistent height)
\coordinate (z4top) at (18.4, \ztop);
\coordinate (z4bot) at (18.4, \zbot);
\begin{scope}[on background layer]
  \node[fit=(query_aug)(vlm_backbone)(angleloss)(z4top)(z4bot),
        fill=violet!6, draw=violet!25, rounded corners=10pt,
        inner sep=10pt] (trainbg) {};
\end{scope}
% Zone 4 caption
\node[zonecaption] at (trainbg.north) {Training Loop};

% =====================================================================
% ARROWS — Catalog image to crop images (object detection)
% =====================================================================

% Catalog → Coffee Table (straight down — symmetric left departure)
\draw[arrowblack] ([xshift=-1.4cm]catalog.south) -- (crop_table.north);

% Catalog → Sofa crop (straight down — aligned centers)
\draw[arrowblack] (catalog.south) -- (crop_sofa.north);

% Catalog → Area Rug (straight down — symmetric right departure)
\draw[arrowblack] ([xshift=1.4cm]catalog.south) -- (crop_rug.north);

% =====================================================================
% ARROWS — Query generation
% =====================================================================

% Sofa crop + "Sectional Sofa" label → Retriever (green Y-fork merge)
% Green tail exits slightly above center of crop_sofa.east
\coordinate (green_merge) at ($([yshift=4pt]crop_sofa.east) + (0.7, 0)$);
\draw[green!50!black, line width=1.2pt] ([yshift=4pt]crop_sofa.east) -- (green_merge);
\draw[green!50!black, line width=1.2pt] (label_sofa.east) -| (green_merge);
\draw[arrowstyle] (green_merge) -- (retriever.west |- green_merge);

% Sofa label → Modifier Generation (straight down — both at x=2.2)
\draw[arrowblack] (label_sofa.south) -- (modifier_gen.north);

% Modifier Generation → modifier text (straight down — both at x=2.2)
\draw[arrowblack] (modifier_gen.south) -- (modifier_text.north);

% Modifier text + Sofa crop → Retriever (red Y-fork merge) — modifier query color
% Red tail exits slightly below center of crop_sofa.east (mirroring the green tail)
\coordinate (red_merge) at (3.5, 1.6);
\draw[red!55!black, line width=1.2pt] (modifier_text.east) -- (red_merge);
\draw[red!55!black, line width=1.2pt] ([yshift=-4pt]crop_sofa.east) -| (red_merge);
\draw[arrowstylemod] (red_merge) -- (retriever.west |- red_merge);

% ANN Index → Retriever (downward)
\draw[arrowblack] (ann_index.south) -- (retriever.north);

% =====================================================================
% ARROWS — Retriever to Retrieved items
% =====================================================================

% Rows 1–2: "Sectional Sofa" similarity query (green) — shared tail, Y-fork
\coordinate (green_fork) at (6.7, 7.6);
\draw[green!50!black, line width=1.2pt] (retriever.east |- green_fork) -- (green_fork);
\draw[arrowstyle] (green_fork) |- (ret_img1.west);
\draw[arrowstyle] (green_fork) |- (ret_img2.west);
% Rows 3–4: "Sofa in grey color" modifier query (red) — shared tail, Y-fork
\coordinate (red_fork) at (6.7, 3.6);
\draw[red!55!black, line width=1.2pt] (retriever.east |- red_fork) -- (red_fork);
\draw[arrowstylemod] (red_fork) |- (ret_img3.west);
\draw[arrowstylemod] (red_fork) |- (ret_img4.west);

% =====================================================================
% ARROWS — Retrieved items to VLM Judge
% =====================================================================

% Rows 1–2: green
\draw[arrowstyle] (ret_img1.east) -- (vlm1.west);
\draw[arrowstyle] (ret_img2.east) -- (vlm2.west);
% Rows 3–4: red
\draw[arrowstylemod] (ret_img3.east) -- (vlm3.west);
\draw[arrowstylemod] (ret_img4.east) -- (vlm4.west);

% =====================================================================
% ARROWS — VLM Judge to Relevance labels
% =====================================================================

% Rows 1–2: green
\draw[arrowstyle] (vlm1.east) -- (label_exact1.west);
\draw[arrowstyle] (vlm2.east) -- (label_nearexact.west);
% Rows 3–4: red
\draw[arrowstylemod] (vlm3.east) -- (label_exact2.west);
\draw[arrowstylemod] (vlm4.east) -- (label_partial.west);

% =====================================================================
% ARROWS — All labels converge to Graded Labeled Pairs (bus topology)
% =====================================================================

% Shared vertical spine x-coordinate
\pgfmathsetmacro{\spinex}{14.8}

% Horizontal branches from each label to the spine (no arrowhead)
\draw[black!70, line width=1.2pt] (label_exact1.east) -- (\spinex, 8.5);
\draw[black!70, line width=1.2pt] (label_nearexact.east) -- (\spinex, 6.5);
\draw[black!70, line width=1.2pt] (label_exact2.east) -- (\spinex, 4.5);
\draw[black!70, line width=1.2pt] (label_partial.east) -- (\spinex, 2.5);

% Vertical spine connecting all branch points (no arrowhead)
\draw[black!70, line width=1.2pt] (\spinex, 8.5) -- (\spinex, 2.5);

% Single arrow from spine to Graded Labeled Pairs (exits at graded_db's y)
\draw[arrowblack] (\spinex, 5.2) -- (graded_db.west);

% =====================================================================
% ARROWS — Graded Labeled Pairs to Training blocks
% =====================================================================

\draw[arrowblack] (graded_db.east) -- (trainbg.west);

% =====================================================================
% ARROWS — Training blocks to Output
% =====================================================================

\draw[arrowblackthick] (trainbg.east) -- (output_model.west);

% =====================================================================
% FEEDBACK LOOP — from Training zone (4) back to Candidate Retrieval zone (2)
% =====================================================================

% Arrow exits the bottom of the training zone, goes down, then left
% under the entire diagram, then up into the candidate retrieval zone.
\coordinate (fb_bottom_right) at ($(trainbg.south) + (0, -0.8)$);
\coordinate (fb_bottom_left) at ($(retrievalbg.south) + (0, -0.8)$);

\draw[arrowblackthick]
  (trainbg.south) -- (fb_bottom_right)
  -- node[below=2pt, font=\sffamily\small\itshape, text=black!70] {Iterative hard negative mining} (fb_bottom_left)
  -- (retrievalbg.south);

\end{tikzpicture}
}% end resizebox
\caption{End-to-end architecture of GradCIR.}
\label{fig:end_to_end_arch}
\end{figure*}

\section{Related Work}\label{sec:related_work}
\subsection{Visual Search and Multimodal Embeddings}
Production visual-search systems at e-commerce scale~\cite{hadi2015buy,du2022amazon,zhu2024bringing,bell2020groknet,yang2017visual,zhai2019learning,shankar2017deep} typically formulate retrieval as image-similarity matching, learning an embedding space over (query image, catalog image) pairs labelled as same-product or not. Two limitations carry over to all of them: relevance is binary, and the query is image-only. Text-aware variants such as VL-CLIP~\cite{giahi2025vl} and the four-tower image--text formulation of Zhu et~al.~\cite{zhu2024bringing} loosen the latter by also accepting text input in the query, but still treat relevance as binary.

The underlying embedding models split into two families that we use as baselines and compare in our architecture ablation (\S\ref{sec:ablation_arch}). \emph{Late-fusion} encoders such as CLIP~\cite{radford2021learning}, ALIGN~\cite{jia2021scaling}, and SigLIP~\cite{tschannen2025siglip}, VL-CLIP~\cite{giahi2025vl} align image and text in a shared space via a contrastive loss and combine them at inference by fusing (averaging in most cases) unimodal embeddings. This two-tower design is fast to serve but cannot represent ``image is X but text says change attribute Y'' interactions that modifier queries require. \emph{Early-fusion} VLMs such as PaliGemma2~\cite{steiner2024paligemma} (our backbone), Qwen-VL~\cite{wang2024qwen2,bai2025qwen3}, LLaVa~\cite{liu2023visual}, Gemini~\cite{comanici2025gemini}, and ModernVBERT~\cite{teiletche2025modernvbert} jointly process image and text tokens through a transformer, giving richer cross-modal interactions at a higher serving cost. Recent early-fusion retrievers -- GME~\cite{zhang2024gme}, MM-Embed~\cite{lin2024mmembed}, mmE5~\cite{chen2025mme5}, Qwen3-VL-Embedding~\cite{qwen3vlembedding} are trained on synthetic (query, item) pairs from web image--caption corpora but, like the industrial systems above, supervise with binary positive/negative labels. GradCIR is, to our knowledge, the first to train an early-fusion VLM retriever on graded relevance for composed multimodal retrieval.

\subsection{Composed Image Retrieval}\label{sec:related_cir}
Composed image retrieval (CIR)~\cite{du2025survey}, also referred to as composed multimodal retrieval in recent surveys, formulates the query as an image plus a textual modifier, and asks the model to retrieve an item that satisfies both. \emph{Classical CIR} approaches train an image--text fusion module on top of a frozen vision--language encoder. ARTEMIS~\cite{delmas2022artemis}, PL4CIR~\cite{zhao2022pl4cir}, TG-CIR~\cite{wen2023tgcir}, CLIP4CIR~\cite{baldrati2022clip4cir}, BLIP4CIR \cite{liu2024bi}, and SPRC~\cite{bai2024sentence} are some of the representative methods in this family. SPN4CIR~\cite{feng2024spn4cir} extends SPRC by mining additional positive and negative triplets, and is the current strongest peer-reviewed supervised baseline on FashionIQ that we compare against.

A second line of work scales CIR with \emph{synthetic triplet supervision}: MagicLens~\cite{zhang2024magiclens} mines 36.7M open-ended instruction triplets from the web; CompoDiff~\cite{gu2023compodiff} generates triplets with latent diffusion; CoVR~\cite{ventura2024covr} mines triplets from web video captions; CoLLM~\cite{huynh2025collm} couples an LLM with CLIP/BLIP backbones and trains on a 3.4M-triplet synthetic dataset (MTCIR). These approaches produce strong zero-shot and supervised CIR models but treat the target as a single binary positive and require an explicit triplet-construction step.

A third line targets \emph{zero-shot CIR} by mapping the visual modality into the text space of a frozen CLIP encoder: Pic2Word~\cite{saito2023pic2word}, SEARLE~\cite{baldrati2023zero}, LinCIR~\cite{gu2024language}, Context-I2W~\cite{tang2024context}, CIReVL~\cite{karthik2024vision} (which uses GPT-4 at inference time), and Slerp-TAT~\cite{jang2024spherical} are representative methods in this family. These methods avoid CIR-triplet training but inherit the binary-target evaluation protocol.

\textbf{Graded relevance is the gap.} None of the above CIR works, to our knowledge, trains on ordinal graded relevance. Standard CIR datasets (FashionIQ \cite{wu2021fashioniq}, CIRR \cite{liu2021image}, CIRCO \cite{baldrati2023zero}) themselves encode a single positive target per query. In production retrieval over a real catalog this is a poor fit: many candidates partially satisfy a query, and the ordering of those partial matches is what determines the customer experience. GradCIR explicitly targets this regime by training on VLM-assigned 4-level graded labels rather than binary triplets.

\section{Methodology}\label{sec:methodology}
\subsection{Problem Formulation and Notation}\label{sec:formulation}
Let $\mathcal{P} = \{p_i = (v_i, w_i)\}_{i=1}^{N_c}$ denote the product catalog, where $v_i$ and $w_i$ are the image and title of item $p_i$ respectively, and $N_c$ is the total number of items in the catalog. A user query $q = (v^q, w^q)$ is a pair of an image and text. The graded-relevance composed multimodal retrieval problem is to learn an encoder $f_\theta$ producing $d$-dimensional embeddings $e_q = f_\theta(q)$ and $e_p = f_\theta(p)$ such that the ranking induced by $\cos(e_q, e_p)$ over $\mathcal{P}$ respects a 4-level ordinal relevance label $r \in \{0, 1, 2, 3\}$, corresponding to \textsc{Irrelevant} / \textsc{PartialMatch} / \textsc{NearExact} / \textsc{Exact}.

\subsection{Data Preparation}\label{sec:data_preparation}
Using large language models to generate high-quality training data has become a popular approach in recent years~\cite{wang2024improving,chen2024bge}. We follow a similar approach with vision--language models (VLMs) and curate a graded-relevance dataset from Walmart catalog. Figure~\ref{fig:end_to_end_arch} shows the end-to-end architecture of GradCIR.

\textbf{Query generation.} For each catalog item, we apply a VLM-based object detector to identify objects from product types of interest, and crop one image per detected product. The cropped image together with the detected object name forms a \emph{similarity query}. A rich catalog image can sometimes contain very large number of products, so we limit ourselves to at most $m{=}10$ similarity queries per catalog item. To support diverse queries, we also generate \emph{modifier queries} by prompting the same VLM to produce modifier text that alters one or more attributes of the cropped product. Using the same VLM for object detection and modifier-text generation saves compute and ensures the generated modifiers are visually consistent with the query image. We use Gemini-2.5-pro \cite{comanici2025gemini} as the VLM for both object detection and modifier-text generation. The final query set $\mathcal{Q}$ contains both query types -- similarity and modifier queries.

\textbf{Candidate retrieval and labeling.} For each $q \in \mathcal{Q}$, we retrieve top-$K$ candidate items $C_q = \{c_1, \dots, c_K\} \subset \mathcal{P}$ using a pretrained multimodal embedding model. A strong VLM judge -- Gemini-2.5-pro then labels every pair: $r(q, c_k) = \phi(q, c_k) \in \{$ \textsc{Exact}, \textsc{NearExact}, \textsc{PartialMatch}, \textsc{Irrelevant} $\}$ (see Figure~\ref{fig:graded_relevance} for representative examples), using the verbatim prompt detailed in Appendix~\ref{appx:prompt}.

\textbf{Query augmentation.} To improve the model's robustness to real-world query distributions, we apply several augmentations per labeled pair: using the uncropped whole image instead of the crop, removing the query text, replacing the object name with a longer object description, or replacing the object name with only the modified attributes. Each augmentation produces a new (query, item) pair sharing the original relevance label. We remove the redundant augmentations, where cropped image is same as original image or object description is same as object name while expanding the dataset. Table~\ref{tab:data_augmentation} enumerates the augmentation matrix.

\textbf{Iterative hard negative mining.} The pretrained retriever used to mine $C_q$ has a bias towards the kind of partial matches it is good at. Relying on it alone leaves the resulting dataset thin on the hard negatives that matter most for a deployed retriever. We therefore run an iterative relevance-feedback loop. Let $\mathcal{D}_t$ denote the labeled dataset at round $t$ (including augmentations) and $f^{(t)}_\theta$ the retriever trained on $\mathcal{D}_t$. At round $t{+}1$, for each query we mine an additional candidate set with $f^{(t)}_\theta$, deduplicate against previous rounds, label the new pairs with the same VLM judge, and form $\mathcal{D}_{t+1} = \mathcal{D}_t \cup \mathcal{D}_t^{\text{new}}$. We stop when offline NDCG@10 on a held-out query set plateaus. In practice the loop converges in three rounds and produces a final dataset of ${\sim}$3.5M labeled (query, item) pairs, of which ${\sim}$2.1M are similarity queries and ${\sim}$1.4M are modifier queries. The final training set is
$$
\mathcal{D} = \{(q, p, \phi(q, p)) \mid q \in \mathcal{Q}, p \in C_q \cup C_q^{HN}\}
$$
where $C_q^{HN}$ is the set of hard negatives mined from the in-training retriever for query $q$.

\begin{table}[t]
    \centering
    % \scriptsize
    % \setlength{\tabcolsep}{3pt}
    % \renewcommand{\arraystretch}{1.1}
    \caption{Query augmentation matrix.}
    \begin{tabular}{c|l|cc}
        \hline
        \multirow{2}{*}{\textbf{Type}} & \multirow{2}{*}{\textbf{Query Text}} & \multicolumn{2}{c}{\textbf{Query Image}}\\
        \cline{3-4}
        & & \textbf{Crop} & \textbf{Whole}\\
        \hline
        \multirow{3}{*}{Similarity}
        & Blank & \checkmark & \checkmark\\ \cline{2-4}
        & Object name & \checkmark & \checkmark\\ \cline{2-4}
        & Object description & \checkmark & \textcolor{red}{\ding{55}}\\
        \hline
        \multirow{2}{*}{Modifier}
        & Modified attributes & \checkmark & \checkmark\\ \cline{2-4}
        & Name + mod.\ attributes & \checkmark & \checkmark\\
        \hline
    \end{tabular}
    \label{tab:data_augmentation}
\end{table}

\subsection{Model Architecture}\label{sec:architecture}
The retriever is a Siamese bi-encoder: a single early-fusion VLM with shared weights encodes queries and items into a common embedding space. Among compact early-fusion options -- PaliGemma2 \cite{steiner2024paligemma}, SmolVLM \cite{marafioti2025smolvlm}, T5Gemma 2 \cite{zhang2025t5gemma}, Granite Vision \cite{team2025granite} we choose PaliGemma2 for its favourable quality/latency tradeoff at ${\sim}$3B parameters.

\textbf{Backbone.}
PaliGemma2 combines a SigLIP-400M vision encoder with a Gemma~2 2.6B language model, connected via a linear projection that maps patch embeddings into the language model's token space. Images are encoded at $224{\times}224$ resolution into $256$ patch tokens, text is tokenized with SentencePiece and capped at $128$ tokens. The interleaved image-patch and text tokens are processed jointly by the Gemma~2 decoder stack, yielding a contextualized early-fusion representation at every position.

\textbf{Retrieval adaptation.}
Since PaliGemma2 is a generative VLM with a prefix-LM attention pattern (bidirectional over the image and prompt prefix, causal over the generated suffix) and does not natively expose fixed-length embeddings, we adapt it for retrieval in three ways:
\begin{enumerate}
    \item \emph{Decoding disabled.} We remove the language-modelling head entirely; the model is used purely as an encoder.
    \item \emph{Last-token pooling.} The hidden state at the last non-padding position serves as the sequence embedding. Because we treat the entire (image, text) input as the bidirectional prefix (no suffix is generated at encoding time), every token, including the final one, attends to all image and text tokens, so its hidden state aggregates the full multimodal context without requiring an additional pooling layer or a dedicated \texttt{[CLS]} token.
    \item \emph{Role-specific prefixes.} A short prefix token --- \texttt{[QUERY]} or \texttt{[ITEM]} --- is prepended to the input text, allowing the shared encoder to learn asymmetric representations for the two roles.
\end{enumerate}

We train with Matryoshka representation learning~\cite{kusupati2022matryoshka}, applying nested losses at dimensions $\{128, 256, 512, 1024\}$, so that embeddings can be truncated to a lower dimensionality at deployment time to meet latency constraints without retraining (see \S\ref{sec:deployment}). For a query $q = (v^q, w^q)$ and a candidate item $p = (v^p, w^p)$, encoder $f_\theta$ produces embeddings
$$
e_q = f_\theta(v^q, w^q), \qquad e_p = f_\theta(v^p, w^p),
$$
with the relevance score $s(q, p) = \cos(e_q, e_p)$.

\subsection{Training Objective}\label{sec:loss}
Bi-encoder retrievers are commonly trained with multiple-negatives ranking loss (MNRL)~\cite{henderson2017efficient} or InfoNCE, which both rely on in-batch negatives and treat relevance as binary: every non-positive item in the batch is an equally bad negative. This is a poor fit for graded relevance, where a \textsc{NearExact} item and an \textsc{Irrelevant} item carry very different penalties for being ranked equally with the positive.

We instead use AngleLoss~\cite{li2023angle,li2024aoe}, which transforms graded labels into angular constraints in the embedding space. For two training pairs $(q_a, p_a, r_a)$ and $(q_b, p_b, r_b)$ with $r_a > r_b$, AngleLoss enforces $\Delta\theta_{q_a,p_a} < \Delta\theta_{q_b,p_b}$, where $\Delta\theta_{q,p}$ denotes the angle between $e_q$ and $e_p$. The loss is
\begin{equation}\label{eq:angle}
    \mathcal{L}_{\text{angle}} = \log \!\left[ 1 + \sum_{(a,b):\, r_a > r_b} \exp\!\left( \frac{\Delta\theta_{q_a,p_a} - \Delta\theta_{q_b,p_b}}{\tau}\right) \right],
\end{equation}
summed over all ordered pairs in the batch with strictly greater relevance, where $\tau$ is a temperature hyperparameter. Because the constraint set in Eq.~\ref{eq:angle} is enumerated over \emph{every pair of distinct grade levels} rather than positive-vs-negative only, the loss exploits all ordering constraints implied by the graded relevance scale ( \textsc{Exact}$>$\textsc{NearExact}, \textsc{Exact}$>$\textsc{PartialMatch}, \textsc{Exact}$>$\textsc{Irrelevant}, \textsc{NearExact}$>$\textsc{PartialMatch}, \textsc{NearExact}$>$\textsc{Irrelevant}, \textsc{PartialMatch}$>$\textsc{Irrelevant} ). Binary contrastive losses collapse five of these six constraints into one, which is the mechanism we ablate in \S\ref{sec:ablation_graded}.

\section{Experiments and Results}\label{sec:experiments}
\subsection{Experimental Setup}
\subsubsection{Implementation Details}
We train our primary GradCIR instantiation -- GradCIR-PG2 on the 3.5M-pair graded dataset $\mathcal{D}$ described in \S\ref{sec:data_preparation}. The model is trained for 2 epochs with AdamW optimizer (learning rate $1{\times}10^{-5}$) with batch size 512 using LoRA~\cite{hu2022lora} (rank 64, $\alpha{=}64$, dropout=0.1 applied to all attention and MLP layers) on top of a PaliGemma2 backbone. The objective is AngleLoss (Eq.~\ref{eq:angle}) with $\tau{=}0.05$. We learn 1024-dimensional Matryoshka embeddings~\cite{kusupati2022matryoshka} with nested losses at 128, 256, 512 and 1024 for post-hoc dimensionality reduction in production. Training runs on 4$\times$NVIDIA~H100-80GB with mixed-precision and gradient checkpointing.

\subsubsection{Baselines}
We compare GradCIR-PG2 against four contemporary multimodal embedding baselines: FashionCLIP~\cite{chia2022contrastive}, SigLIP2~\cite{tschannen2025siglip}, GME-Qwen~\cite{zhang2024gme}, and Qwen3-VL-Embedding~\cite{qwen3vlembedding}. All baselines are evaluated both off-the-shelf and after fine-tuning with the GradCIR recipe (same 3.5M graded dataset, identical AngleLoss + LoRA setup); the fine-tuned variants are denoted \textbf{+GradCIR} in the tables. For late-fusion baselines (FashionCLIP, SigLIP2) we use the average of image and text embeddings for pre-trained models and weighted average (with learned weights) for fine-tuned models. For early-fusion baselines (GME-Qwen, Qwen3-VL-Embedding) we use the native fused embedding.

\subsubsection{Evaluation Datasets}\label{sec:eval_datasets}

\begin{table}[t]
    \centering
    \small
    \caption{Overview of the WVST dataset (Ex.{=}Exact, NE{=}NearExact, PM{=}PartialMatch, Irr.{=}Irrelevant).}
    \begin{tabular}{c|c|c|c|c|c|c}
        \hline
        \multirow{2}{*}{\textbf{Query Type}} & \textbf{\#} & \textbf{\#Total} & \multicolumn{4}{c}{\textbf{Candidates per Query}}\\
        \cline{4-7}
        & \textbf{Queries} & \textbf{Cands.} & \textbf{Ex.} & \textbf{NE} & \textbf{PM} & \textbf{Irr.}\\
        \hline
        \multicolumn{7}{c}{\textbf{Home Vertical}} \\
        \hline
        Similarity & 1073 & 216K & 9.55 & 21.50 & 128.43 & 42.82 \\
        Modifier & 1140 & 231K & 4.57 & 25.31 & 120.29 & 52.17 \\
        \hline
        \multicolumn{7}{c}{\textbf{Fashion Vertical}} \\
        \hline
        Similarity & 843 & 170K & 6.03 & 18.23 & 118.94 & 62.59 \\
        Modifier & 854 & 171K & 4.35 & 25.12 & 112.97 & 60.69 \\
        \hline
    \end{tabular}
    \label{tab:eval_label_distribution}
\end{table}

To evaluate the performance of our proposed methodology, we construct a high-quality internal test set, the Walmart Visual Search Test (WVST), and also report results on two standard public benchmarks for fashion retrieval: Street2Shop~\cite{hadi2015buy} and FashionIQ~\cite{wu2021fashioniq}.

\textbf{Walmart Visual Search Test (WVST)}. A high-quality internal test set spanning home and fashion verticals, constructed by stratified sampling of catalog products and user-uploaded review images, cropped with Gemini-2.5-pro. For each query, we construct a labeled candidate pool by taking the union of top-$N$ retrievals from multiple embedding models, then labeling each candidate with one of the four graded-relevance levels using a Gemini-2.5-pro judge. We report results separately for similarity and modifier queries, for the home and fashion verticals. The candidate-label distribution is shown in Table~\ref{tab:eval_label_distribution}.

\textbf{Street2Shop}. A similarity-query benchmark: given a street photo of a clothing item, retrieve the matching shop product. The test split contains 6770 street-image queries and 9802 shop-image candidates. We use the street image plus its category label (e.g., ``dress'', ``top'') as the query and the shop image plus category label as the candidate.

\textbf{FashionIQ}. A modifier-query benchmark: given a reference image and a natural-language modifier (e.g., ``make it striped''), retrieve the target item. We evaluate on the validation split following the standard protocol of averaging Recall@10 and Recall@50 across the Dress, Shirt, and Top-tee sub-categories. We use FashionIQ rather than CIRR/CIRCO because the latter draw from NLVR2/COCO natural-scene imagery that is out-of-distribution for a catalog-trained model; we discuss this scope choice further in \S\ref{sec:scope_limitations}.

\subsubsection{Evaluation Metrics}
For WVST we report two graded-relevance metrics: \textbf{Adjusted Recall@5}~(Adj-R@5) and \textbf{NDCG@\{5,10\}}, both computed over the labeled candidate pool with NDCG gain values 3 / 2 / 1 / 0 for the Exact / NearExact / PartialMatch / Irrelevant levels respectively. We report Adj-R@5 separately for \emph{Exact} (Exact + NearExact only) and \emph{Partial} (Exact + NearExact + PartialMatch). Adjusted recall normalizes by $\min(K, n_{\text{rel}})$ rather than $K$, to fairly handle queries with fewer than $K$ relevant items. $n_{\text{rel}}$ is the number of relevant items in the candidate pool for the query.
\begin{equation*}
\text{Adj-R}@K = \frac{|\text{relevant items in top-}K|}{\min(K,\, n_{\text{rel}})}.
\end{equation*}
For Street2Shop we report Recall@\{1,5,10\} following standard practice. For FashionIQ we report Recall@\{10, 50\} averaged across sub-categories.

\subsection{Results on Walmart Visual Search Test}\label{sec:results_wvst}

\begin{table}[t]
    \centering
    \scriptsize
    \setlength{\tabcolsep}{4pt}
    \renewcommand{\arraystretch}{1.05}
    \caption{Performance on WVST. ``\textbf{+GradCIR}'' denotes fine-tuning with our graded-relevance recipe.}
    \begin{tabular}{l|cc|cc||cc|cc}
        \hline
        \multirow{3}{*}{\textbf{Model}} & \multicolumn{4}{c||}{\textbf{Home Vertical}} & \multicolumn{4}{c}{\textbf{Fashion Vertical}} \\
        \cline{2-9}
            & \multicolumn{2}{c|}{\textbf{Adj-R@5}} & \multicolumn{2}{c||}{\textbf{NDCG}} & \multicolumn{2}{c|}{\textbf{Adj-R@5}} & \multicolumn{2}{c}{\textbf{NDCG}} \\
            & Exact & Partial & @5 & @10 & Exact & Partial & @5 & @10 \\
        \hline
        \multicolumn{9}{c}{\textbf{Similarity Queries}} \\
        \hline
        FashionCLIP                    & 0.3507 & 0.8837 & 0.7694 & 0.7740 & 0.3992 & 0.9153 & 0.8178 & 0.8213 \\
        ~~~~\textbf{+GradCIR}             & 0.4474 & 0.9225 & 0.8267 & 0.8281 & 0.4346 & 0.9266 & 0.8395 & 0.8446 \\
        SigLIP2                        & 0.4820 & 0.9004 & 0.8217 & 0.8174 & 0.5156 & 0.9416 & 0.8681 & 0.8630 \\
        ~~~~\textbf{+GradCIR}             & 0.5525 & 0.9133 & 0.8467 & 0.8429 & 0.5549 & 0.9405 & 0.8761 & 0.8713 \\
        GME-Qwen                       & 0.4049 & 0.9044 & 0.8030 & 0.8066 & 0.4075 & 0.9214 & 0.8228 & 0.8289 \\
        ~~~~\textbf{+GradCIR}             & 0.5806 & 0.9442 & 0.8770 & 0.8752 & 0.5954 & 0.9524 & 0.9000 & 0.8985 \\
        Qwen3-VL-Emb                   & 0.5522 & 0.9359 & 0.8602 & 0.8555 & 0.5956 & 0.9530 & 0.8922 & 0.8883 \\
        ~~~~\textbf{+GradCIR}             & 0.6406 & \textbf{0.9597} & 0.9012 & 0.8984 & 0.6620 & 0.9720 & 0.9203 & 0.9147 \\
        \textbf{GradCIR-PG2}          & \textbf{0.6560} & 0.9588 & \textbf{0.9066} & \textbf{0.9034} & \textbf{0.6747} & \textbf{0.9721} & \textbf{0.9241} & \textbf{0.9216} \\
        \hline
        \multicolumn{9}{c}{\textbf{Modifier Queries}} \\
        \hline
        FashionCLIP                    & 0.3619 & 0.8545 & 0.7674 & 0.7712 & 0.4031 & 0.8826 & 0.8018 & 0.8035 \\
        ~~~~\textbf{+GradCIR}             & 0.4496 & 0.8939 & 0.8181 & 0.8218 & 0.4642 & 0.9000 & 0.8313 & 0.8339 \\
        SigLIP2                        & 0.3167 & 0.8734 & 0.7687 & 0.7721 & 0.3860 & 0.9132 & 0.8198 & 0.8234 \\
        ~~~~\textbf{+GradCIR}             & 0.4460 & 0.8840 & 0.8025 & 0.8085 & 0.4866 & 0.9151 & 0.8459 & 0.8486 \\
        GME-Qwen                       & 0.4152 & 0.8775 & 0.8001 & 0.8006 & 0.4213 & 0.8926 & 0.8191 & 0.8217 \\
        ~~~~\textbf{+GradCIR}             & 0.4998 & 0.9133 & 0.8488 & 0.8517 & 0.5153 & 0.9162 & 0.8588 & 0.8613 \\
        Qwen3-VL-Emb                   & 0.3909 & 0.9067 & 0.8110 & 0.8118 & 0.4627 & 0.9170 & 0.8421 & 0.8429 \\
        ~~~~\textbf{+GradCIR}             & 0.5470 & 0.9320 & 0.8677 & 0.8687 & 0.5752 & 0.9377 & 0.8846 & 0.8859 \\
        \textbf{GradCIR-PG2}          & \textbf{0.5704} & \textbf{0.9401} & \textbf{0.8810} & \textbf{0.8807} & \textbf{0.5908} & \textbf{0.9401} & \textbf{0.8930} & \textbf{0.8942} \\
        \hline
    \end{tabular}%
    \label{tab:results_wvst}
\end{table}

As shown in Table~\ref{tab:results_wvst}, GradCIR-PG2 achieves the best performance on WVST, both in Adj-R and NDCG. It improves over the next-best variant, Qwen3-VL-Emb(+GradCIR), by an average of 2.83\% for Adj-R@5 (Exact) and 0.91\% for NDCG@10 (average is taken over all query types and for both Home and Fashion categories). The gap is small on similarity queries and widens on modifier queries, where graded supervision matters most.

A second observation addresses whether the gains come from the data or the backbone. Every baseline benefits substantially from the GradCIR recipe -- FashionCLIP, SigLIP2, GME-Qwen, and Qwen3-VL-Embedding gain 24.3\% of Adj-R@5 (Exact) on average over their off-the-shelf counterparts, validating the data-pipeline contribution independently of backbone choice. The remaining gap between GradCIR-fine-tuned baselines and GradCIR-PG2 reflects the backbone's suitability for graded ranking.

\subsection{Results on Street2Shop}\label{sec:results_street2shop}

\begin{table}[t]
    \centering
    \small
    \setlength{\tabcolsep}{4pt}
    \renewcommand{\arraystretch}{1.1}
    \caption{Retrieval performance on the Street2Shop test split.}
    \vspace{-0.1cm}
    \begin{tabular}{l|ccc}
        \hline
        \textbf{Model} & \textbf{R@1} & \textbf{R@5} & \textbf{R@10}\\
        \hline
        FashionCLIP            & 0.3121 & 0.5084 & 0.5932 \\
        ~~~~\textbf{+GradCIR}     & 0.3329 & 0.5230 & 0.5975 \\
        SigLIP2                & 0.5180 & 0.6739 & 0.7357 \\
        ~~~~\textbf{+GradCIR}     & 0.5558 & 0.7064 & 0.7718 \\
        Qwen3-VL-Emb           & 0.6443 & 0.7718 & 0.8227 \\
        ~~~~\textbf{+GradCIR}     & \textbf{0.6631} & \textbf{0.7886} & \textbf{0.8304} \\
        \textbf{GradCIR-PG2}  & 0.6464 & 0.7815 & 0.8266 \\
        \hline
    \end{tabular}
    \vspace{-0.1cm}
    \label{tab:results_street2shop}
\end{table}

Table~\ref{tab:results_street2shop} reports the retrieval performance on Street2Shop dataset. GradCIR-PG2 performs closely to Qwen3-VL-Emb(+GradCIR) (within 0.4--1.7 points across all $K$), and improves over other baseline models by large margins. This confirms that the graded-relevance training transfers well to the binary-target Street2Shop evaluation protocol, and that our training strategy consistently improves performance for similarity queries.

\subsection{Results on FashionIQ}\label{sec:results_fashioniq}

\begin{table}[t]
    \centering
    \small
    \setlength{\tabcolsep}{4pt}
    \caption{FashionIQ zero-shot CIR Performance.}
    \vspace{-0.1cm}
    \begin{tabular}{l|c|ccc}
        \hline
        \textbf{Model} & \textbf{Backbone} & \textbf{R@10} & \textbf{R@50} & \textbf{Avg}\\
        \hline
        Pic2Word~\cite{saito2023pic2word}            & CLIP-L/14 & 0.2470 & 0.4370 & 0.3420\\
        SEARLE~\cite{baldrati2023zero}             & CLIP-L/14 & 0.2560 & 0.4620 & 0.3590\\
        LinCIR~\cite{gu2024language}                   & CLIP-L/14 & 0.2640 & 0.4660 & 0.3650\\
        Context-I2W~\cite{tang2024context}           & CLIP-L/14 & 0.2780 & 0.4890 & 0.3835\\
        Slerp-TAT~\cite{jang2024spherical}               & CLIP-L/14 & 0.2830 & 0.4760 & 0.3795\\
        CIReVL (GPT-4)~\cite{karthik2024vision}      & CLIP-L/14 & 0.2860 & 0.4860 & 0.3860\\
        CoLLM~\cite{huynh2025collm}                  & CLIP-L/14 & 0.3010 & 0.4950 & 0.3980\\
        Slerp-TAT~\cite{jang2024spherical}               & BLIP-L/16 & 0.3280 & 0.5330 & 0.4305\\
        CoLLM~\cite{huynh2025collm}                  & BLIP-L/16 & \textbf{0.3460} & \textbf{0.5600} & \textbf{0.4530}\\
        \textbf{GradCIR (zero-shot)} & PaliGemma2 & 0.3036 & 0.4964 & 0.4000\\
        \hline
    \end{tabular}
    \vspace{-0.1cm}
    \label{tab:fashioniq_zeroshot}
\end{table}

\begin{table}[t]
    \centering
    \small
    \setlength{\tabcolsep}{4pt}
    \caption{FashionIQ supervised CIR Performance (fine-tuned on the FashionIQ training split).}
    \vspace{-0.1cm}
    \begin{tabular}{l|c|ccc}
        \hline
        \textbf{Model} & \textbf{Backbone} & \textbf{R@10} & \textbf{R@50} & \textbf{Avg}\\
        \hline
        ARTEMIS~\cite{delmas2022artemis}             & w/o VLP       & 0.2605 & 0.5029 & 0.3817 \\
        ComqueryFormer~\cite{xu2023multi}   & w/o VLP       & 0.2937 & 0.5536 & 0.4237 \\
        PL4CIR~\cite{zhao2022pl4cir}                 & CLIP          & 0.3902 & 0.6300 & 0.5101 \\
        TG-CIR~\cite{wen2023tgcir}                   & CLIP-ViT-B/16 & 0.3981 & 0.6306 & 0.5144 \\
        CLIP4CIR~\cite{baldrati2022clip4cir}         & CLIP-RN50$\times$4 & 0.4252 & 0.6560 & 0.5406 \\
        BLIP4CIR~\cite{liu2024bi}              & BLIP-base     & 0.4349 & 0.6731 & 0.5540 \\
        SPRC~\cite{bai2024sentence}                      & BLIP-2        & 0.5492 & 0.7497 & 0.6485 \\
        SPN4CIR~\cite{feng2024spn4cir}               & BLIP-2        & 0.5637 & 0.7645 & 0.6641 \\
        \textbf{GradCIR (fine-tuned)} & PaliGemma2 & \textbf{0.5684} & \textbf{0.7721} & \textbf{0.6703}\\
        \hline
    \end{tabular}
    \vspace{-0.1cm}
    \label{tab:fashioniq_supervised}
\end{table}

We evaluate GradCIR-PG2 on FashionIQ in two settings: \emph{zero-shot} (Walmart-trained model used as-is) and \emph{fine-tuned} (additionally fine-tuned on the FashionIQ training split with $r{=}64$ LoRA for 5 epochs).

\textbf{Zero-shot.} Without any fashion-CIR fine-tuning, GradCIR (Table~\ref{tab:fashioniq_zeroshot}) reaches average Recall of 0.40, matching or exceeding every published CLIP-L-class zero-shot CIR method. Larger, more recent, BLIP-L variants (CoLLM-BLIP-L, Slerp-TAT-BLIP-L) remain ahead. The results demonstrate that catalog-only graded supervision transfers competitively to the binary-CIR evaluation protocol, with no triplet construction step required.

\textbf{Supervised.} With FashionIQ fine-tuning (Table~\ref{tab:fashioniq_supervised}), GradCIR reaches average recall of 0.6703, ahead of the strongest peer-reviewed supervised baseline we compare against, SPN4CIR (0.6641), reaching the state-of-the-art among the supervised CIR methods on FashionIQ.
\subsection{Ablations}\label{sec:ablations}

\subsubsection{Graded vs.\ Binary Supervision}\label{sec:ablation_graded}

\begin{table}[t]
    \centering
    \small
    \caption{Effect of label granularity on retrieval quality.}
    \begin{tabular}{l|cc|cc}
        \hline
        \multirow{2}{*}{\textbf{Supervision}} & \multicolumn{2}{c|}{\textbf{Sim. Queries}} & \multicolumn{2}{c}{\textbf{Mod. Queries}} \\
        \cline{2-5}
            & NDCG@5 & NDCG@10 & NDCG@5 & NDCG@10 \\
        \hline
        Binary (2-level)               & 0.8762 & 0.8698 & 0.8370 & 0.8376 \\
        3-level                        & 0.8775 & 0.8783 & 0.8412 & 0.8450 \\
        \textbf{4-level} & \textbf{0.9154} & \textbf{0.9125} & \textbf{0.8870} & \textbf{0.8875} \\
        \hline
    \end{tabular}
    \label{tab:ablation_graded}
\end{table}

Table~\ref{tab:ablation_graded} isolates the effect of label granularity: all rows share the same backbone (PaliGemma2) and training data (3.5M pairs); only the number of relevance levels changes. NDCG is averaged across Home and Fashion. Moving from binary to 4-level supervision yields an improvement of 4.5\%--6\% in NDCG@5 and 4.9\%--5.9\% in NDCG@10 across both query types. The intermediate 3-level setting (which merges NearExact and PartialMatch into a single grade) already improves over binary, but a substantial gap to 4-level remains. This confirms that distinguishing NearExact from PartialMatch---precisely the discrimination that graded relevance adds---is essential for top-of-ranking quality.

\subsubsection{Effect of Modifier-Query Training}\label{sec:ablation_modifier}

\begin{table}[t]
    \centering
    \small
    \caption{Effect of including modifier queries in training.}
    \begin{tabular}{l|cc|cc}
        \hline
        \multirow{2}{*}{\textbf{Training Data}} & \multicolumn{2}{c|}{\textbf{Sim.\ Adj-R@5}} & \multicolumn{2}{c}{\textbf{Mod.\ Adj-R@5}} \\
        \cline{2-5}
            & Exact & Partial & Exact & Partial \\
        \hline
        Similarity only            & 0.6258 & 0.9530 & 0.3016 & 0.9004 \\
        \textbf{Sim.\ + Modifier} & \textbf{0.6654} & \textbf{0.9655} & \textbf{0.5806} & \textbf{0.9401} \\
        \hline
    \end{tabular}
    \label{tab:ablation_modifier}
\end{table}

Table~\ref{tab:ablation_modifier} compares training on similarity queries alone vs.\ training on both query types. Adj-R@5 is averaged across Home and Fashion verticals. We observe that training the model on modifier queries in addition to similarity queries makes the model more robust, and improves its performance on similarity queries as well. This is because modifier queries provide additional supervision on fine-grained image--text interactions that are also relevant for similarity-query understanding. Including modifier queries improves similarity-query Adj-R@5 by 6.3\% for exact match retrieval and 1.3\% for partial match retrieval, and modifier-query Adj-R@5 by a much larger 92.5\% for exact match and 4.4\% for partial match. This validates the importance of including modifier queries in training to achieve strong performance on both query types.

\subsubsection{Architecture: Late vs. Early Fusion}\label{sec:ablation_arch}
The fine-tuned-baseline rows of Table~\ref{tab:results_wvst} double as an architecture ablation. On similarity queries, the late-fusion SigLIP2 closes most of the gap to early-fusion baselines after fine-tuning. On modifier queries, it does improve with fine-tuning, but still remains well below the early-fusion baselines. Similarly, FashionCLIP lags behind even after fine-tuning, and the gap to early-fusion models is larger than for SigLIP2. This is consistent with earlier observations (\cite{saito2023pic2word, huynh2025collm}) that late fusion struggles to model the complex image--text interactions required for modifier-query understanding, even with strong fine-tuning.

\subsection{Deployment}\label{sec:deployment}

GradCIR-PG2 is deployed as the retrieval backbone of Walmart's visual-search system, currently serving a partial traffic slice. This is the first production visual-search system at Walmart.

\textbf{Serving architecture.}
At inference time, every catalog item is pre-indexed by its GradCIR-PG2 embedding truncated to 256 dimensions as described in \S\ref{sec:architecture}. User queries are encoded online by the same model and retrieved via approximate nearest-neighbour (ANN) search. The existing text-based reranker is applied on top of the retrieved candidates to produce the final ranking. The full pipeline operates within the production latency SLA.

\section{Scope and Limitations}\label{sec:scope_limitations}
GradCIR targets composed image retrieval \emph{over a structured product catalog} where every candidate is a single sellable item. Two consequences shape our evaluation choices and scope:

\textbf{Why not evaluate on CIRR or CIRCO?}
The CIRR and CIRCO benchmarks draw query and candidate images from NLVR2 and COCO respectively, consisting of natural scenes with multiple objects, and free-form composition. A retriever trained on catalog images is fundamentally out-of-distribution on these benchmarks, and the resulting numbers measure adaptation effort more than retrieval quality. FashionIQ and Street2Shop, by contrast, draw from a fashion catalog and align with our training distribution, which is why we report these as our public benchmarks.

\textbf{What we do not solve.}
We do not address (i) multi-item compositional queries (``find a sofa like this and a rug like that''), which require a more expressive query representation than a single image--text pair; or (ii) explicit attribute-level controllability at inference time, which would benefit from explicit attribute conditioning. Extending the graded-relevance formulation to these settings is left to future work.

\textbf{Reproducibility.}
All architecture choices, hyperparameters, prompts, augmentation details, and recipe details required to reproduce the methodology on any comparable proprietary catalog are provided in the paper. The FashionIQ and Street2Shop results use public data and standard public evaluation protocols.

\section{Conclusion}\label{sec:conclusion}
We presented GradCIR, a graded-relevance methodology for composed multimodal retrieval in e-commerce visual search. The central methodological contribution is the use of 4-level ordinal relevance labels, assigned at scale by a VLM judge and exploited by a hierarchy-aware angular objective. The graded-vs-binary ablation isolates this single design choice and shows that finer relevance granularity is what closes the ranking-quality gap on graded NDCG metrics. The full pipeline curates 3.5M labeled pairs without manual annotation through an iterative relevance-feedback loop. On the internal benchmarks, GradCIR outperforms all baselines on almost every metric, with the largest gains on the more fine-grained graded NDCG metrics. On public benchmarks, GradCIR is competitive with the strongest baselines. The system is deployed in production at Walmart, where it's serving live visual-search user traffic. Future work includes extending the graded formulation to multi-item compositional queries and exploring attribute-level conditioning for more fine-grained relevance modeling.

%%
%% Appendix.
%%
\appendix
\section{Labeling Prompt}\label{appx:prompt}
The verbatim prompt used by the VLM judge $\phi(\cdot,\cdot)$ to assign relevance labels is reproduced below.

\noindent
\begin{tcolorbox}[width=\columnwidth, colupper=lightblack, boxsep=2pt, left=3pt, right=3pt, top=2pt, bottom=2pt, breakable]
\begin{Verbatim}[
    fontsize=\scriptsize,
    breaklines=true,
    obeytabs=true,
    tabsize=2,
    breaksymbol={},
    commandchars=\\\{\}
]
\textbf{\color{black}>>SYSTEM}
You are an expert in product search and relevance evaluation for an e-commerce company. You are meticulous, detail-oriented, and have a deep understanding of user intent in an online shopping context.

Your task is to evaluate the relevance of an e-commerce product (candidate) to a visual search query. The query includes a product image (which may be a partial or obstructed view) and a query text. The query text can explain or override few attributes of the product, the non-overridden attributes(visual or otherwise) remain part of the query. The candidate includes a product image and some product details such as title, description or structured attributes (e.g., size, brand, color).

Step-by-step Instructions:
1. Extract the specifications (attributes, features, and all other important aspects) from both the query and candidate.
Examples include:
  * Intended use, product type, color, shape, size, material, style, brand, intended age, gender, theme, occasion, etc.
  * Cardinal attributes related to usability such as: number of drawers, shelves, windows, doors, parts, pieces, etc.

2. Compare the query and candidate based on:
  * Visual similarity (overall appearance, shape, size, style).
  * Specification alignment.
  * Whether the candidate would fulfill the intent behind the query.
  * Focus on the e-commerce product aspects, not on the lighting, angle, background etc.

3. Assign one of the following relevance labels to the candidate:
  1. Exact: Matches the query exactly in terms of all specifications and visual appearance.
  2. NearExact: Completely satisfies the intent behind the query, but may contain minor difference in appearance or specification (e.g., different brand or slightly varied style or different shade of same color or slanted vs straight legs of a coffee table).
  3. PartialMatch: A candidate with same product type and intended use. Attributes such as color, style, size, material, etc. can be different from the query.
  4. Irrelevant: Either Product type or intended use is different from the query, or candidate is intended for a different age group or gender or size.

Examples:
  * Query: Product A in red color
      Candidate: Variant of product A in crimson color
      Label: NearExact
  * Query: Product A
      Candidate: Very similar to product A with minor visual differences but additional features
      Label: NearExact
  * Query: Product A in red color
      Candidate: Variant of product A with only difference being green color
      Label: PartialMatch

Final Output Format:
  Explanation: <Brief explanation under 100 words>
  Label: <Exact | NearExact | PartialMatch | Irrelevant>

\textbf{\color{black}>>USER}
Query:
  Image: <query image> <query text>
Candidate:
  Image: <candidate product image> <candidate product title>
\end{Verbatim}
\end{tcolorbox}

% \section{Example Retrieval Results}\label{appx:examples}
% \begin{figure*}[t]
%     \centering
%     \includegraphics[width=0.5\textwidth]{../figures/sample_results_orig_p1}
%     \includegraphics[width=0.5\textwidth]{../figures/sample_results_orig_p2}
%     \caption{Sample retrieval results for visual queries without modifier text using GradCIR. The first column shows the query image and corresponding text; the following four columns show the top-4 retrieved catalog items. These illustrate the model's ability to align user-uploaded imagery to studio catalog images despite domain shift.}
%     \label{fig:results_original}
% \end{figure*}

% \begin{figure*}[t]
%     \centering
%     \includegraphics[width=0.5\textwidth]{../figures/sample_results_mod_p1}
%     \includegraphics[width=0.5\textwidth]{../figures/sample_results_mod_p2}
%     \caption{Sample retrieval results for modifier queries using GradCIR. The first column shows the query image and modifier text; the following four columns show the top-4 retrieved catalog items that reflect the requested attribute change.}
%     \label{fig:results_modifier_appx}
% \end{figure*}

% %%
% %% GenAI Usage Disclosure --- unlimited length per CIKM 2026 CFP;
% %% does not count against the 7-page main-body budget.
% %%
% \section*{GenAI Usage Disclosure}\label{sec:genai}
% \input{sections/genai_disclosure}

\printbibliography

\end{document}